| | |
|---|---|
| Title | Synthesis and texturization processes of (super)-hydrophobic fluorinated surfaces by atmospheric plasma |
| Authors | J. Hubert[a], J. Mertens[a], T. Dufour[a], N. Vandencasteele[a], F. Reniers[a], P. Viville[b], R. Lazzaroni[b], M. Raes[c], H. Terryn[c] |
| Affiliations | [a] Faculté des Sciences, Service de Chimie Analytique et de Chimie des Interfaces, Université Libre de Bruxelles, CP-255, Bld du Triomphe, B-1050 Bruxelles, Belgium<br>[b] Service de Chimie des Matériaux Nouveaux, Université de Mons-UMONS/Materia Nova, 20 Place du Parc, 7000 Mons, Belgium<br>[c] Department of Metallurgy, Electrochemistry and Materials Science (SURF), Vrije Universiteit Brussel (VUB), Pleinlaan 2, B-1050 Brussel, Belgium |
| Ref. | Journal of Materials Research, 2015, Vol. 30, Issue 21, 3177-3191 |
| DOI | http://dx.doi.org/10.1557/jmr.2015.279 |
| Abstract | The synthesis and texturization processes of fluorinated surfaces by means of atmospheric plasma are investigated and presented through an integrated study of both the plasma phase and the resulting material surface. Three methods enhancing the surface hydrophobicity up to the production of super-hydrophobic surfaces are evaluated: (i) the modification of a polytetrafluoroethylene (PTFE) surface, (ii) the plasma deposition of fluorinated coatings and (iii) the incorporation of nanoparticles into those fluorinated films. In all the approaches, the nature of the plasma gas appears to be a crucial parameter for the desired property. Although a higher etching of the PTFE surface can be obtained with a pure helium plasma, the texturization can only be created if $O_2$ is added to the plasma, which simultaneously decreases the total etching. The deposition of $C_xF_y$ films by a dielectric barrier discharge leads to hydrophobic coatings with water contact angles (WCAs) of 115°, but only the filamentary argon discharge induces higher WCAs. Finally, nanoparticles were deposited under the fluorinated layer to increase the surface roughness and therefore produce super-hydrophobic hybrid coatings characterized by the nonadherence of the water droplet at the surface. |

# 1. Introduction

Super-hydrophobic thin films are attractive in applications where water-repellent, anti-fouling, and self-cleaning properties are desirable. A super-hydrophobic surface is one that repels water to such an extent that the droplet is almost spherical and easily rolls across the surface. They are generally characterized by a high water contact angle (WCA) (>150°) and a low tilting angle or a low contact angle hysteresis (<10°); the hysteresis being defined by the difference between the advancing contact angle and the receding contact angle.[1] The highest known WCA of a smooth low-energy surface is comprised between 110° and 120° depending on the chemical group present at the surface ($CH_3$, $CF_2$, $CF_3$).[2] Super-hydrophobic surfaces are therefore obtained by combining rough surface morphology and low surface energy coatings.[2] Such coatings have been deposited using wet chemistry with low surface energy materials but these methods usually require multiple processing steps and utilize solvents.[3–6] The use of plasmas is then a very promising synthetic route to produce super-hydrophobic surfaces since this approach has the advantage of reducing the number of steps required to modify the surface of materials. Moreover, the use of atmospheric plasmas presents many benefits such as reducing the treatment time as well as the costs related to high-vacuum systems. Working under atmospheric pressure can therefore be advantageous in industrial applications since the process can be easily implemented in a continuous production line. The two main pathways involving the plasma processing of polymers are (i) the direct modification of polymer surfaces and (ii) the plasma deposition of thin polymeric films.

The plasma treatments of fluoropolymer surfaces have drawn a special interest and particularly in the case of the polytetrafluoroethylene (PTFE) because of its outstanding properties such as high thermal stability, low friction coefficient, hydrophobicity, and chemical inertness which explain its use for biocompatibility and self-cleaning applications.[7] The treatment of PTFE by plasma has been investigated in terms of both hydrophobic[8–10] and hydrophilic[11–14] surfaces. Inert gases such as helium and argon improve the wettability of the PTFE by decreasing its WCA through either a defluorination of the polymer surface or an incorporation of residual oxygen.[15–19] Moreover, it has been shown that, at low pressure, those gases could induce a sputtering of the PTFE and produce a film onto a substrate made of fluorocarbon species ejected from the polymer.[20–23]

The treatment of PTFE by $O_2$-containing gases is more controversial as both hydrophobic and hydrophilic modifications have been obtained. For instance, some studies showed that oxygen radio-frequency (RF) plasmas could lead to an etching of the surface characterized by oxygen grafting and as a result a decrease in WCA.[13,14,24] Salapare et al. observed that in a $O_2$ low-pressure plasma distinct behaviors (hydrophobic or hydrophilic) were obtained depending on the plasma energy.[25] Many other studies also demonstrated that an oxygen plasma treatment could induce a roughening of the samples responsible for the increase in WCA but with no change in the chemical composition.[8–10] Ryan and Badyal observed that the oxygen plasma did not induce any variation in the chemical composition but its use led to a surface texturization.[9] More recently, Vandencasteele et al. highlighted the synergetic role of charged species (electrons) and atomic oxygen in the etching of PTFE by low-pressure oxygen plasmas.[10] Super-hydrophobic PTFE surfaces were also created by an Ar+$O_2$ low-pressure plasma.[26] Although no change in the chemical composition was recorded, leaf-like micro-protrusions were observed after a 4 h-plasma treatment related to the WCA of 158° and characterized by a root mean square (RMS) roughness of about 2 µm. Most of the studies were realized at low pressure; only few researches showed results focused on the superhydrophobicity of PTFE by $O_2$ in atmospheric plasmas. For instance, the treatment of PTFE by the post-discharge of an Ar-$O_2$ plasma torch induced an increase in the WCA (130°) and in the roughness while it was not observed in the pure argon plasma.[27] Trigwell et al. showed that PTFE was chemically resistant to an atmospheric pressure glow discharge supplied with He-$O_2$ with a WCA up to 125°.[28]

Low-pressure plasma deposition of fluorocarbon films has been widely studied in the last decades.[29–34] Gaseous monomers (e.g., $CF_4$, $C_2F_4$, $C_2F_6$, $C_3F_8$, $C_3HF_7$ or c-$C_4F_8$) polymerized into the discharge led to the deposition of coatings with different properties depending on the plasma source or the feed gases. The $F/CF_x$ ratio was identified as one of the most relevant internal indicators to describe the deposition of the fluoropolymers because it defines the competition between the etching and the polymerization processes.[35–37] The precursors used at atmospheric pressure are usually identical to those used in low-pressure processes but the use of a plasma gas (usually helium or argon) is most often required when working at atmospheric pressure. The groups of d'Agostino[30,33,35,38,39] and Vinogradov[40,41] have been very active in this domain at both low and atmospheric pressures. For instance, they widely studied the influence of reactive gases such as $H_2$ and $O_2$ on the chemistry of the films and the deposition rate. They showed that the addition of oxygen tends to consume the $CF_2$ through the formation of COF, CO, $CO_2$, and/or F, then reducing the deposition rate and shifting the process from deposition to etching. Regarding the wettability properties, only few works reported in the literature succeeded in getting WCAs higher than 100–110°, at atmospheric pressure.[42–44]

As previously mentioned, the synergy of the roughness and the low-energy surface is required to enhance the hydrophobicity. Many studies have investigated the combination of those two factors to create super-hydrophobic surfaces where authors combine several techniques; from creating roughness into a material followed by an hydrophobization step to obtain the desired properties. For instance, photolithography was used to transfer patterns to silicon wafers before hydrophobization by silanization reagents,[3] electrochemical deposition of zinc oxide, nickel, and copper leads to hydrophobic surfaces after modification with hydrophobic self-assembly monolayers such as fluoroalkylsilane,[45,46] an oxygen plasma etching of paper fibers followed by the deposition of a pentafluoroethane thin film.[47] Additionally to those methods, the roughness might be obtained by the incorporation of nanoparticles which ones can be deposited on the substrate by solvent evaporation,[48] sol–gel method,[49,50] electrophoretic path,[51] or spray-deposition methods.[52,53]

Once the nanoparticles are deposited, the hydrophobization step is performed through the synthesis of a low surface energy film such as fluorinebased compounds. Few studies have reported the interest of making hybrid coatings from nanoparticles; such as the use of a sol–gel process in the presence of tetraethoxysilane,54 the use of a vacuum-based method combining the deposition of nanoparticles by means of a gas aggregation source with conventional RF magnetron sputtering of PTFE,55 or also the incorporation of hydrophobically modified silica nanoparticles in PTFE emulsion.56

In the present study, we show that cold atmospheric plasma is an efficient technology to easily obtain (super)-hydrophobic surfaces from fluorinated compounds. Three pathways are described: (i) the modification of an existing fluoropolymer surface, (ii) the deposition of a fluorocarbon film and (iii) the enhancement of the film hydrophobicity through the addition of nanoparticles. Based on the integrated study of both the plasma phase and the resulting material surface, polymerization and texturization processes are proposed. The influence of the gas phase on the etching of the PTFE and the deposition and texturization of $C_xF_y$ compounds are described. Although helium and argon have both demonstrated their efficiency in the plasma deposition of fluorocarbons, studies usually deal with the effects of a single gas, either argon or helium. Moreover, these last results are compared with super-hydrophobic hybrid coatings made of nanoparticles covered by a fluorinated layer deposited by a dielectric barrier discharge (DBD).

## 2. Experimental part

### 2.1. Materials

One mm thick PTFE samples were supplied by Goodfellow. The samples were cleaned in pure methanol (AnalaR Normapur, VWR) and pure iso-octane (GR for analysis, Merck), before being exposed to the plasma postdischarge. The liquid precursors namely perfluoro-2-methyl-2-pentene ($C_6F_{12}$) and perfluorohexane ($C_6F_{14}$) were provided by Fluorochem and were used without any further purification (Fig. 1). 80 nm $SiO_2$ nanoparticles from Alfa Aesar are dispersed in demineralized water without any surfactant. Silicon wafers (100) from Compart Technology Ltd. (Tamworth, UK) were used as substrates after being cleaned with methanol and isooctane. The glow discharge was sustained with argon (Air Liquide, ALPHAGAZ 1), helium (Air Liquide, ALPHAGAZ 1), or $O_2$ (Air Liquide, ALPHAGAZ 1).

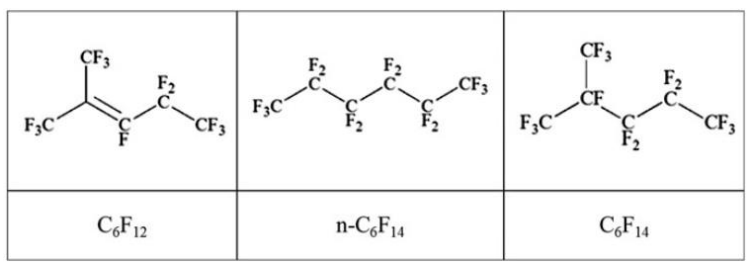

FIG. 1. Structures of the two liquid precursors $C_6F_{12}$ and $C_6F_{14}$ and the two $C_6F_{14}$ isomers.

### 2.2. The plasma sources

The PTFE sample surfaces were exposed to the linear post-discharge of an RF atmospheric plasma torch, the Atomflo™ 400L-Series from SurfX Technologies (Redondo Beach, California), supplied with He and He-$O_2$ and integrated to a robotic system detailed in a previous paper.57 The kinematic parameters used in this study are the following ones: power 90–120 W, $\Phi_{He}$ 15 L/min, $\Phi_{O_2}$ 0–0.1 L/min, while the kinematic parameters are the scanning length ($L_S$ = 10 mm), the scanning velocity ($v_S$ = 25 mm/s), the number of scans ($N_S$ = 1000), and the torch-to-sample distance ($g_S$ = 0.5–1 mm). In all our experiments, the scanning plasma source moved back and forth along a single axis. According to the used kinematic parameters ($L_S$ = 10 mm, $v_S$ = 25 mm/s), the treatment time is the number of scans multiplied by the time required to scan one scanning length (0.4 s). The width of the plasma torch slit being 0.8 mm, the exposure time of a spot (0.8 mm) is estimated at 0.032 s. For a number of scans of 1000, the total treatment time is 400 s while the exposure time of each spot is about 32 s. The plasma-polymerized (pp)-fluorinated films were deposited with a home-built DBD described in previous studies.[58,59] The vapors of the precursor were introduced into the discharge by means of the carrier gas (second flow—0.2 L/min) bubbling into the liquid, after being diluted with a primary flow of the carrier gas with a total carrier gas flow of 5 L/min. The injected amounts for 0.2 L/min are: Ar–$C_6F_{12}$ 782 6 44 mg/min, He–$C_6F_{12}$ 783 6 52 mg/min, Ar–$C_6F_{14}$ 768 6 84 mg/min, and He–$C_6F_{14}$ 767 6 101 mg/min. The operating frequency was set at 17.1 kHz with an output power set at 50 W for the results presented in this paper, supplied by an AFS-G10S power generator (AFS Entwicklungs-und Vertriebs GmbH, Horgau, Germany). The deposition time varied from 30 to 360 s.

## 2.3. Water contact angles (WCA)

A drop shape analyzer (Krüss DSA 100, Krüss GmbH, Hamburg, Germany) was used to measure dynamic WCAs onto the samples, according to the sessile drop method. The fitting method used is the "Tangent 1". Advancing and receding contact angles were both measured by growing and shrinking the size of a single drop on the surface sample, from 0 to 15 µL and back to 0 µL at a rate of 30 µL/min. The receding angles are not presented in this paper but information can be found in Refs. 59, 60.

## 2.4. X-ray photoelectron spectroscopy (XPS)

XPS analysis was performed on a Physical Electronics PHI-5600 photoelectron spectrometer (Eden Prairie, Minnesota). Survey spectra were used to determine the elemental chemical composition of the surface. Narrow region photoelectron spectra were used for the chemical study of the C1s. The spectra were acquired using the Mg anode (1253.6 eV) operating at 300 W. Wide surveys were acquired at a pass-energy of 187.5 eV with a five-scans accumulation (time/step: 50 ms, eV/step: 0.8) and high-resolution spectra of the C1s peaks were recorded at a pass-energy of 23.5 eV with an accumulation of 5 scans (time/step: 200 ms, eV/step: 0.05). The elementary composition was calculated after the removal of a Shirley background and using the sensitivity coefficients from the manufacturer's handbook: $S_C$ = 0.205, $S_F$ = 1, and $S_O$ = 0.63.

## 2.5. Atomic force microscopy (AFM)

AFM was used to analyze the surface morphology of the deposited films. AFM images were recorded in air with a Nanoscope IIIa microscope (Brüker, Karlsruhe, Germany) operating in tapping mode. The probes were commercially available silicon tips with a spring constant of 24–52 N/m, a resonance frequency lying in the 264–339 kHz range, and a typical radius of curvature in the 5–10 nm range. The images presented here are height images recorded with a sampling resolution of 512*512 data points and a scan size of 5*5 µm$^2$.

## 2.6. Profilometry

The thickness measurements were performed using a stylus profiler Brücker dektak XT (Brüker, Karlsruhe, Germany). The stylus with a 2 µm radius scans the surface with a force of 1 mg and the measurement was controlled and analyzed with the Vision 64 software. The thickness was estimated based on the height difference between the substrate surface and the surface of the coating and used to estimate the deposition rates. The roughness of the samples containing nanoparticles was estimated by a 3D simulation of the surface obtained by the sum of 150 scans covering a surface of 0.9 mm$^2$. The RRMS was then calculated by the software. The force of the stylus for these measurements was set at 0.2 mg.

## 2.7. Scanning electron microscopy (SEM)

SEM measurements were performed on a JEOL JSM-7000F (JEOL, Tokyo, Japan) equipped by a Schottky field emission gun. The analyses were essentially topographic with a magnification of 5000. The electrons were accelerated by a 5 kV electrical field.

## 2.8. Mass loss measurements and XPS of aluminum trapping ejected species

The PTFE samples were weighted before and after the plasma treatments to evaluate mass variations. A Sartorius BA110S Basic series analytical balance characterized by a capacity of 110 g and a readability of 0.1 mg was used. Moreover, during the plasma treatment, an aluminum foil was placed close to the samples. As aluminum is known to be an efficient fluorine trap,[61] we then analyzed the foil by XPS, looking for the presence of fluorinated species.

## 2.9. Optical emission and absorption spectroscopy (OES–OAS)

OES was performed with a SpectraPro-2500i spectrometer from ACTON Research Corporation (Acton, Massachusetts) (0.500 m focal length, triple grating imaging). The light emitted by the discharge was collected by an optical fiber and transmitted to the entrance slit (50 µm) of the monochromator. Each optical emission spectrum was acquired with the 1800 grooves/mm grating (blazed at 500 nm) and recorded over 10 accumulations with an exposure time of 50 ms for the DBD and 30 accumulations, 25 ms for the postdischarge. To tackle intensity variations as a function of the plasma conditions, the emissions of all the species were divided by the emission of the whole spectra (i.e., a continuum ranging from 250 to 850 nm). Line-absorption spectroscopy was applied to study He ($2^3$S) metastable states, more precisely on the transition $2^3$S–$3^3$P at 388.9 nm.[62] The absorption rate A is related to the emission intensities by the relation A = ($I_L$ + $I_P$ – $I_{L+P}$)/$I_L$, where $I_P$ is the light intensity emitted from the post-discharge, $I_L$ is the line intensity from the external lamp and $I_{L+P}$ is the line intensity from the two light sources.[63,64]

## 2.10. Mass spectrometry (MS)

MS of the gas phase was performed with a Hiden analytical atmospheric gas analysis—quadrupole gas analyzer (QGA; Hiden Analytical, Warrington, UK). The gases were collected through a perfluoroalkoxy capillary located between the electrodes at the boundary of the plasma region. The secondary electron multiplier detector (SED) was used to detect fragments with low partial pressures ($10^{-6}$–$10^{-13}$ Torr). To avoid excessive fragmentation of the precursor in the ionization chamber, the electron energy was set at 35 eV. The provided software MASsoft7 was used either to analyze the partial pressure as a function of the m/z ratio or to follow the partial pressure as a function of time for several specific m/z ratios, simultaneously.

# 3. Results & Discussion

As previously mentioned, three pathways are presented to show the efficiency of the cold atmospheric plasma to easily obtain (super)-hydrophobic surfaces from fluorinated compounds.

## 3.1. PTFE surface properties enhanced by atmospheric post-discharge

### 3.1.1. Characterization of the surface modifications

The first evaluated method to enhance the hydrophobicity of a surface is the use of an atmospheric plasma torch to modify polymer properties. In this study, PTFE surfaces were submitted to the post-discharge of a RF plasma torch supplied in He or in He-$O_2$ to study wettability variations. By comparing pure helium and He-$O_2$ plasma treatments, it is shown that oxygen has a significant impact on the PTFE wettability and its presence is required to enhance the hydrophobicity of the polymer surface. As illustrated in Fig. 2, helium plasma treatments lead to a small decrease in the WCA while an increase in the hydrophobicity is observed by adding oxygen. Moreover, we previously showed that advancing WCA as high as 155° (with a contact angle hysteresis < 10°) could be obtained for higher number of scans and therefore a higher treatment time.[57]

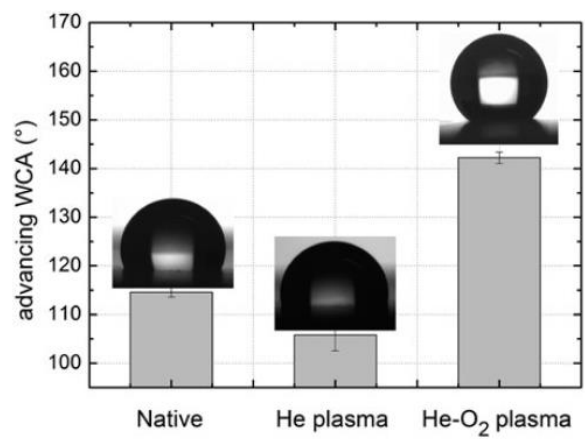

FIG. 2. Advancing WCAs of untreated PTFE and He and He–$O_2$ plasma-treated PTFE, with the associated pictures of the static WCA. (P = 90 W, $L_s$ = 10 mm, $v_s$ = 25 mm/s, $g_s$ = 500 µm, $N_s$ = 1000, $\Phi(He)$ = 15 L/min, $\Phi(O_2)$ = 0 and 100 mL/min).

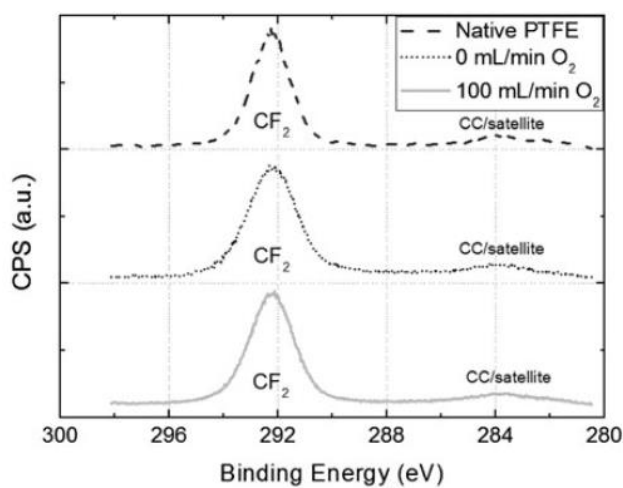

FIG. 3. HR C1s spectra of untreated PTFE and He and He-$O_2$ plasma-treated PTFE. ($L_s$ = 10 mm, $v_s$ = 25 mm/s, $g_s$ = 500 µm, Ns 5 1000, $\Phi(He)$ = 15 L/min, $\Phi(O_2)$ = 0 and 100 mL/min, P = 90 W).

Although modifications in wettability were observed, the relative surface chemical composition measured by XPS was kept unchanged as shown in Fig. 3. Indeed, only a strong $CF_2$ component at 292.2 eV and a small CC component (284.6 eV) mixed up with the satellite peak of $CF_2$ are detected at the surface with the same High Resolution C1s profile independently of the plasma conditions (native, He, He-$O_2$).

It is known that a high roughness and a low-energy surface are required to get (super)-hydrophobic properties. 2,65–67 According to the above results in terms of surface composition (which does not change), the variation in the WCA is assumed to be only induced by morphology modifications. Indeed, the RMS roughness slightly decreases for a pure helium plasma (from 17 to 15 nm), while this roughness is much higher when oxygen is added (up to 58 nm). This significant variation between both plasma treatments is illustrated by the AFM images in Fig. 4 and justifies the wettability behaviors. Moreover, these results are correlated with those previously observed at low pressure.10,68

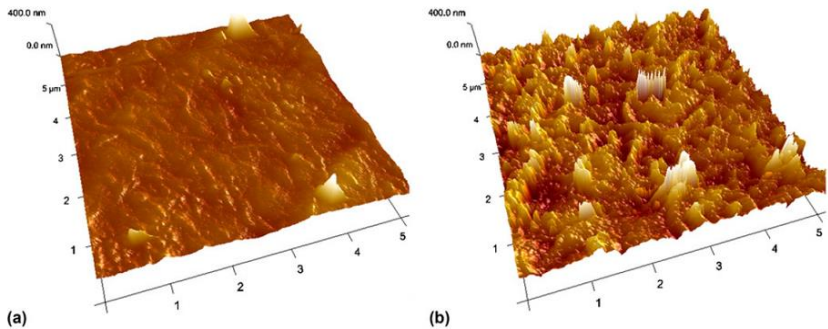

FIG. 4. 5*5 µm$^2$ AFM 3D images of a PTFE surface treated by (a) pure He plasma ($R_{RMS}$ 15 nm) and (b) He-$O_2$ plasma ($R_{RMS}$ 58 nm), with a Z scale of 511 nm. ($L_s$ = 10 mm, $v_s$ = 25 mm/s, $g_s$ = 500 lm, $N_s$ = 1000, $\Phi$(He) = 15 L/min, $\Phi$($O_2$) = 0 and 100 mL/min, P = 90 W).

Although the surface modifications by a pure helium plasma are less noticeable in terms of wettability and morphology, this treatment however appears to induce the highest etching of the PTFE surface. Indeed, higher mass losses and higher concentrations of fluorinated compounds (% F values from survey spectra and % $CF_2$ values from C1s high resolution spectra) were detected onto an aluminum foil, indicating that more species are ejected from the PTFE sample using pure helium treatment as illustrated in Fig. 5 (pure helium plasma corresponds to 0 mL/min $O_2$ flow rate). From the XPS study of the aluminum foil, the ejection of a higher amount of material in the pure He plasma can also be evidenced by the attenuation of the signal coming from the aluminum substrate indicating a thicker deposit of polymer (i.e., $O_2$ 100 mL/min: 8% F, 17% C, 39% O and 36% Al compared to no oxygen: 43% F, 27% C, 12% O and 18% Al).

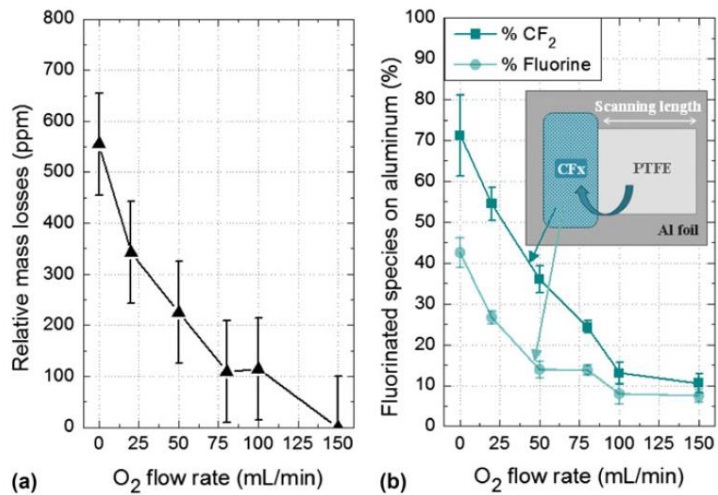

FIG. 5. (a) Relative mass losses of PTFE as a function of the oxygen flow rate and (b) relative surface composition of the fluorinated fragments ejected from the PTFE on the aluminum foil. ($L_s$ = 10 mm, $v_s$ = 25 mm/s, $g_s$ = 1 mm, $N_s$ = 1000, $\Phi$(He) = 15 L/min, P = 90 W).

**3.1.2. Characterization of the post-discharge**

OES–OAS of the post-discharge allowed detecting the emissive species existing in the post-discharge and analyzing their evolution as a function of the oxygen flow rate. Absorption spectroscopy showed that helium metastable atoms were present in the post-discharge and were consumed by the addition of oxygen leading to the formation of atomic oxygen according to the following reactions69:

(i) Consumption of the helium metastable through the Penning Ionization with oxygen: $He^m + O_2 \rightarrow O_2^+ + He + e$

(ii) Production of atomic oxygen: $O_2^+ + e \rightarrow 2O$

The evolution of the most relevant species observed by spectroscopy is illustrated in Fig. 6. The intensity of $O_2^+$ ions (525 nm) permanently increases as a function of the oxygen flow rate. Several mechanisms are usually considered to explain their production [dissociative transfer with $He_2^+$ ions, charge transfer reaction involving $O_2$ molecules and $O^+$ ions or the electron impact ionization of $O_2$ (Ref. 62)]. Considering the species observed or absent from the post-discharge and the weak electron densities measured in typical He-$O_2$ RF post-discharge,70 the mechanism explaining the production of $O_2^+$ ions is assumed to be the Penning ionization of $O_2$ molecules by He metastable atoms [Eq. (1)]. These He metastable species were indeed experimentally evidenced by OAS as seen in Fig. 6(a) and their consumption by oxygen are consistent with the production of $O_2^+$. Moreover, their presence can also be emphasized by the observation of $N_2^+$ ions. Indeed, $N_2$ molecules easily quench $He^m$ to produce $N_2^+$ through the Penning ionization of $N_2$.71,72 Their intensity decreases with the oxygen flow rate since helium metastable atoms are consumed by $O_2$. Finally, the most probable channel to produce atomic oxygen is the electron impact dissociation of $O_2^+$ [Eq. (2)] since the rate constant of this reaction is important compared to other reactions.62 The plateau observed for the highest oxygen flow rate can be interpreted as a result from equilibrium between the production and the consumption mechanisms.

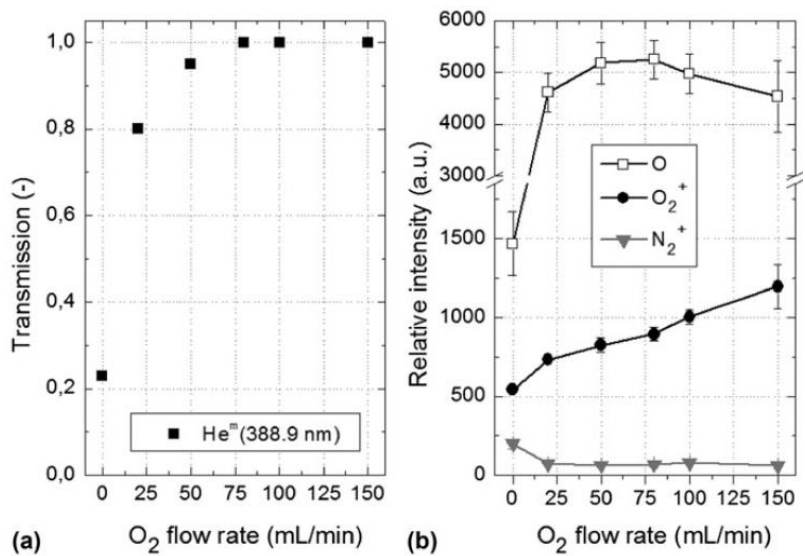

FIG. 6. (a) Transmission rate of He $2^3S$ at 388.9 nm and (b) optical emission intensity of O (844.6 nm), $O_2^+$ (525 nm) and $N_2^+$ (391 nm) in the He-$O_2$ post-discharge ($g_s$ = 1 mm, $\Phi$(He) = 15 L/min, P = 120 W).

According to the results obtained from OES–OAS and surface characterization, etching processes explaining the two different behaviors are proposed. In a He–$O_2$ plasma, no modification of the surface composition was observed but the roughness and therefore the WCA were increased inducing the (super)-hydrophobic property. In this condition, we assume an anisotropic etching where the oxygen atoms etch preferentially the amorphous phase of the polymer. Oxygen atoms are indeed known to etch more easily the amorphous region of a polymer than the crystal phase,40,73 creating alveolar structures at the surface playing a role on the increase in roughness. Since the etching of the PTFE at atmospheric pressure is higher in absence of oxygen but leads to a smoothening of the surface (slight decrease in WCA and $R_{RMS}$), we assume a layer-by-layer random physical etching without any preferential orientation. Studies have shown that helium could be responsible for the sputtering of PTFE.21–23 For instance, the treatment of PTFE by a glow discharge supplied in helium induced the ejection of solid polymer and a mass loss of the polymer.22 The main product identified in the helium plasma-induced decomposition of PTFE was the TFE monomer ($CF_2=CF_2$). Moreover, the RF plasma sputtering of PTFE is considered to involve scission of $-(CF_2)_n-$ chains yielding smaller $-(CF_2)_m-$ (m << n) segments, which are deposited at the substrate surfaces and form a structure dominated by $-(CF_2)-$ groups.74 The scission of $-(CF_2)-$ chains is therefore supposed to be due to the highly energetic helium metastable atoms. Indeed, the helium metastable species, being consumed by the reaction with oxygen through the Penning ionization, are not available any more to etch the PTFE surface.74,75 The increase in the oxygen flow rate leads therefore to a reduction of mass losses and of species detected onto the aluminum foil. To have a better understanding and visualization, the two processes are depicted in a schematic representation in Fig. 7.

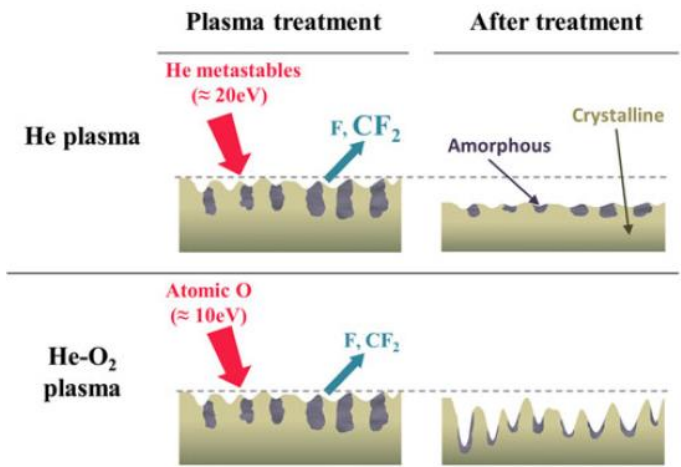

*FIG. 7. Scheme of the two possible etching processes. In a pure helium plasma, the most important fragment produced is $CF_2$.*

## 3.2. Plasma deposition of hydrophobic fluorinated coatings by DBD at atmospheric pressure

**3.2.1. Characterization of the plasma phase**

The second method using plasmas to create (super)-hydrophobic surfaces is the Plasma-Enhanced Chemical Vapor Deposition (PECVD) of precursors containing hydrophobic groups such as $CF_x$ groups which present the lower surface energy. Two liquid precursors, namely, perfluoro-2-methyl-2-pentene ($C_6F_{12}$) and perfluorohexane ($C_6F_{14}$) were used to deposit coatings in argon and helium. The films produced by both precursors have shown similar properties when comparing treatments in argon and helium.

The study of the plasma phase by MS and optical spectroscopy, completed by electrical measurements highlighted the influence of the gas nature and of the discharge regime (filamentary versus glow) on the fragmentation of the precursors.39 Figure 8 presents the evolution of the intensity of some significant m/z fragments detected my MS, in a He (dashed curves) or Ar (full curves) plasma. The energetic streamers present in the filamentary argon discharge induce a higher fragmentation evidenced by the decrease in the main fragments intensity in the mass spectrometer as seen in Fig. 8(a). In helium, the effect of the plasma is less pronounced, but all the intensities of the detected fragments increase when the discharge is turned on, as shown by the dashed curves in Fig. 8(a). $C_6F_{14}$ is held up as an example since all its main fragments are detected in the mass spectrometer range, but a similar behavior has been observed for $C_6F_{12}$.

Despite the complexity of the reactions in the discharges, differences between the two systems ("MS" versus "plasma 1 MS") are briefly illustrated in the following scheme (Fig. 9). As the fragmentation of the precursor is carried out upstream of the mass spectrometer, some of the $C_3F_7^+$, $CF_3^+$, and $C_3F_5^+$ ions are already present in the discharge (especially with an Ar plasma) and would be even more fragmented by the mass spectrometer, explaining their decreasing intensity and the increasing intensity of smaller fragments such as CF or $CF_2$.

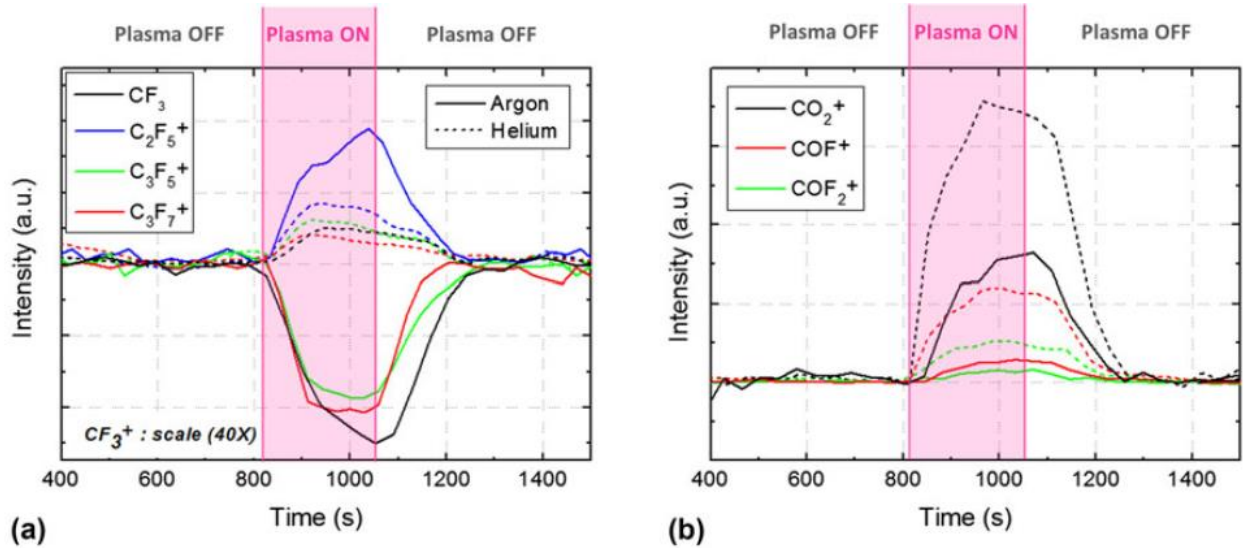

FIG. 8. Intensity of the (a) $CF_3^+$, $C_2F_5^+$, $C_3F_5^+$ and $C_3F_7^+$ and (b) $CO_2^+$, $COF^+$, $COF_2^+$ fragments generated during the atmospheric pressure PECVD of $C_6F_{14}$ as a function of time, in argon (_____ full curves) and helium (- - - dashed curves). (0.2 L/min $C_6F_{14}$-carrier gas, 5 L/min carrier gas total, 50 W).

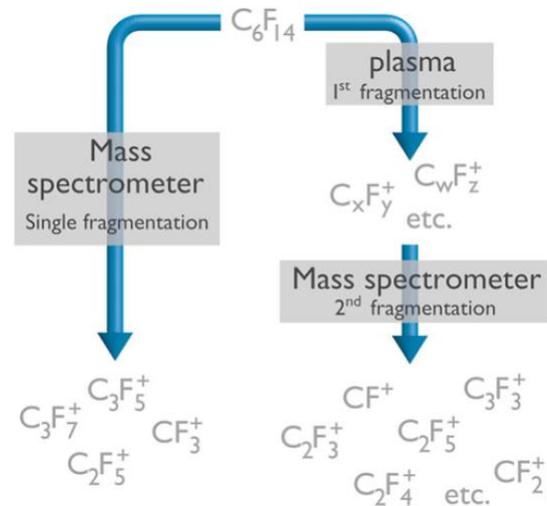

FIG. 9. Scheme of the possible fragmentation of $C_6F_{14}$ (left) in the mass spectrometer and (right) in the plasma, followed by the mass spectrometer ($C_xF_y^+$ and $C_wF_z^+$ assuming to be $CF_3^+$, $C_3F_7^+$ and $C_3F_5^+$ according to MS results).

Additionally to these observations, we highlighted a crucial role of the oxygen impurities in the plasma polymerization. MS results have shown that oxygenated fragments such as $CO_2^+$, $COF^+$, or $COF_2^+$ were highly detected in the helium plasma compared to the argon discharge [Fig. 8(b)]. The efficiency of generating atomic oxygen in helium plasma has also been evidenced by OES.[59] The emission intensity of the atomic oxygen lines (e.g., 777 nm) is indeed strongly increased when $O_2$ is added to the helium discharge while small variations are observed in argon. This behavior towards oxygen can be explained by the excited states of helium that are energetic enough to ionize most atoms and molecules and are therefore very efficient to excite impurities. Indeed, metastables atoms of helium have energies of 19.82 and 20.62 eV while the 11.55 and 11.72 eV argon metastable atoms are too low to induce the Penning ionization of $O_2$ (ionization potential of 12.07 eV[76,77]).

### 3.2.2. Characterization of the film properties

The surface properties reflect the results obtained from the plasma phase. The deposition rates are more than 5 times higher in argon than in helium (i.e., 240 and 25 nm/min for $C_6F_{12}$ in argon and helium, respectively; 25 and 5 nm/min for $C_6F_{14}$). The difference between the two precursors is assumed to be related to the higher reactivity of the C=C double bond present in $C_6F_{12}$, similarly to the case of hydrocarbon or acrylate molecules.[78–80] The surface composition of the films polymerized in both argon and helium estimated from XPS measurements are in good correlation with the behavior of the plasma phase. The $C_6F_{12}$ is held up as an example. The chemical structure of the pp-$C_6F_{12}$ film is better preserved in the case of helium since a large amount of $CF_3$ is detected, while the argon plasma leads to a more disordered XPS spectrum containing a higher concentration of $CF_2$ (Table I).

|  | Surface elementary composition (%) | | | Chemical environment of C1s (%) | | | | |
|---|---|---|---|---|---|---|---|---|
|  | F1s | O1s | C1s | $CF_3$ | $CF_2$ | CF | CCF | CC |
| Ar–$C_6F_{12}$ | 53.9 ± 0.9 | 0 | 46.1 ± 0.9 | 23.4 ± 2.0 | 28.8 ± 1.1 | 22.0 ± 0.3 | 23.7 ± 1.3 | 2.2 ± 0.7 |
| He–$C_6F_{12}$ | 51.4 ± 0.9 | 2.1 ± 0.3 | 46.6 ± 0.4 | 29.5 ± 0.7 | 20.6 ± 0.7 | 18.5 ± 0.4 | 29.3 ± 1.3 | 2.0 ± 0.4 |

TABLE I. Surface composition of pp-$C_6F_{12}$ films as a function of the carrier gas (0.2 L/min $C_6F_{12}$-carrier gas, 5 L/min carrier gas total, 50 W).

Given the structure of the precursor in Fig. 1, as well as the lower electron density, the lower energy and the lower microdischarges density[76,81–83] of the helium discharge, we suggest that the precursor structure is more preserved when using helium. Moreover, a small percentage of oxygen (about 3%) is detected only in the films polymerized in helium. As previously detailed, helium is known to be reactive toward oxygen impurities. [76] This property is important as it can explain why small amounts of oxygen are detected in the film and why the energetic helium species are less involved in the fragmentation/polymerization of the monomer. These species could indeed be consumed by the impurities.

The hydrophobic and morphological properties depend on the nature of the precursor and the gas phase as shown in Fig. 10. Although all the coatings are hydrophobic with a WCA of at least 115°, only the filamentary argon discharge induces a significant roughening of the surface, hence an increase in the hydrophobicity. Indeed, WCAs of 135 and 140° associated to $R_{RMS}$ higher than 40 and 60 nm were obtained for Ar–$C_6F_{12}$ and Ar–$C_6F_{14}$, respectively, while the homogeneous helium discharge leads to a smooth, low-energy surface characterized by a WCA of 115° and a $R_{RMS}$ of about 1 nm, for similar thicknesses.

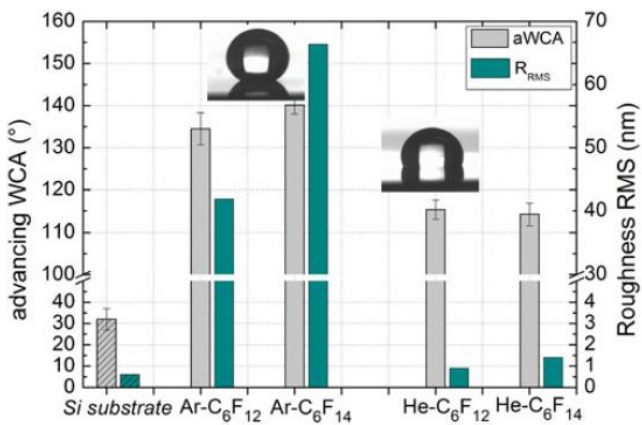

FIG. 10. Advancing WCA and roughness RMS of the silicon wafer substrate and the pp films. Illustration of the drop behavior of (left) Ar–$C_6F_{14}$ and (right) He–$C_6F_{12}$ coatings. (0.2 L/min precursor-carrier gas, 5 L/min carrier gas total, 50 W).

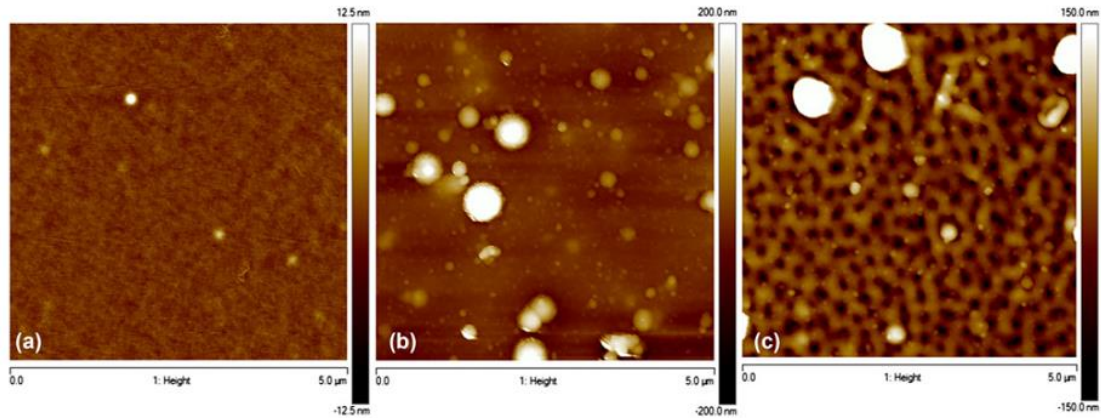

FIG. 11. 5*5 µm² AFM images of (a) He–$C_6F_{12}$, (b) Ar–$C_6F_{12}$ and (c) Ar–$C_6F_{14}$ pp-films. (0.2 L/min $C_6F_{14}$-carrier gas, 5 L/min carrier gas total, 50 W, thickness of about 300 nm).

The AFM images in Fig. 11 illustrate the three typical morphologies of films deposited from He-$C_6F_{12}$, Ar-$C_6F_{12}$ or Ar-$C_6F_{14}$. The first picture [Fig. 11(a)] shows the smooth surface obtained from a He-$C_6F_{12}$ plasma deposition. The second is characteristic of the Ar-$C_6F_{12}$ films and the last one of the Ar–$C_6F_{14}$ films. For a similar thickness, the Ar–$C_6F_{14}$ plasma seems to induce a higher roughness and hydrophobicity as shown in Fig. 10. Moreover, as illustrated in Fig. 11(c), this $C_6F_{14}$ precursor induces a roughness structure totally different from the $C_6F_{12}$ precursor. The alveolar structure of approximately 100 nm are quite regular and could be characteristic of the filamentary discharge inducing a localized etching of the surface in competition with the polymerization. The presence of the C=C double bond in $C_6F_{12}$ and its lower F/C ratio could promote the polymerization character instead of the etching, as it is the case for $C_6F_{14}$. The etching could be responsible for this alveolar structure observed in Ar–$C_6F_{14}$ coatings. Moreover, the alveolar structure could induce the slightly higher WCAs recorded on these coatings.

## 3.3. Effect of nanoparticles on the hydrophobic fluorinated coatings properties

**3.3.1. Effect of nanoparticles: Enhancement of the hydrophobic property**

The results shown before confirm that a high roughness combined with a low-energy surface is required to produce (super)-hydrophobic coatings. The low-energy surface was brought by the fluorinated groups of the coatings but the intrinsic roughness resulting from the filamentary argon discharge was not high enough to reach a super-hydrophobic state, i.e. a WCA higher than 150°. Indeed, the highest advancing WCA obtained was about 140° with a high contact angle hysteresis of 20–30°. Therefore, we developed a new process in which the addition of nanoparticles enhanced the surface texturization. Typical double-scale roughness is usually required to obtain super-hydrophobic surfaces with similar wettability properties to the lotus leaf present in nature84,85 Neinhuis and Barthlott, in 1997, obtained pictures of a lotus leaf as seen where the structures consist of a combination of a two-scale roughness: 10 lm (rough structure) and 100 nm (fine structure).86,87 Additional studies showed that particles with a mean diameter of 60 and 70 nm were efficient to increase the hydrophobicity of the surface.55,88

In this study, nanoparticles were deposited on a silicon substrate through the evaporation of a 0.05 g/L dispersion of 80 nm $SiO_2$ nanoparticles (volume of the solution 1.5 mL for a substrate surface of about 20 cm²). After evaporation of the solvent (water), the substrate was covered by a $C_6F_{12}$ film synthesized by atmospheric plasma using argon or helium as carrier gas. The $C_6F_{12}$ precursor has been chosen for this study because of its higher deposition rate and therefore the more accurate characterization of the deposited films. WCAs were measured to study the influence of the additional roughness on the wettability properties of the film.

As presented in Fig. 12, higher WCAs are observed when nanoparticles are located under the plasma synthesized film. An increase in the contact angle value of about 15° was measured with the nanoparticles when helium was used as carrier gas while only a rise of 8° was measured with argon, for an identical deposition time. Since the deposition rates are different for the two gases (i.e., 240 and 25 nm/min for argon and helium, respectively), we also report the WCA values for coatings with a similar thickness. From Fig. 12, we show that the argon plasma leads to higher WCAs as the presence of nanoparticles with helium plasma reaches a maximum WCA of about 130°.

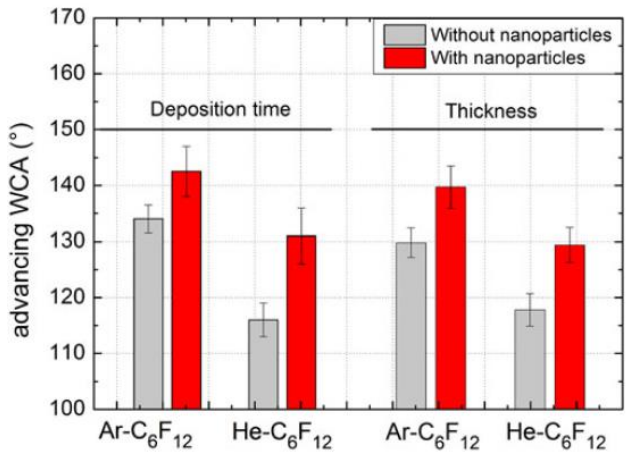

FIG. 12. Advancing WCA of the pp films without nanoparticles (light gray) and with nanoparticles (red) for an identical deposition time of 2 min (left) and a similar thickness of about 200 nm (right). (0.2 L/min $C_6F_{12}$-carrier gas, 5 L/min carrier gas total, 50 W).

To discriminate between the two factors responsible for those wettability properties (surface roughness and the coating chemical composition), we carried out XPS measurements. As shown in the C1s high resolution spectra represented in Fig. 13, the presence of nanoparticles has no significant influence on the chemical properties of the upper layer of the plasma synthesized film. This absence of chemical variations has been observed regardless of the carrier gas (argon or helium), and could indicate that the main part of the polymerization occurs in the plasma phase and does not depend on the substrate and/or that the $SiO_2$ nanoparticles are not reactive towards the fluorinated precursor. According to those results, the increase in WCAs does not result from a change in chemical composition.

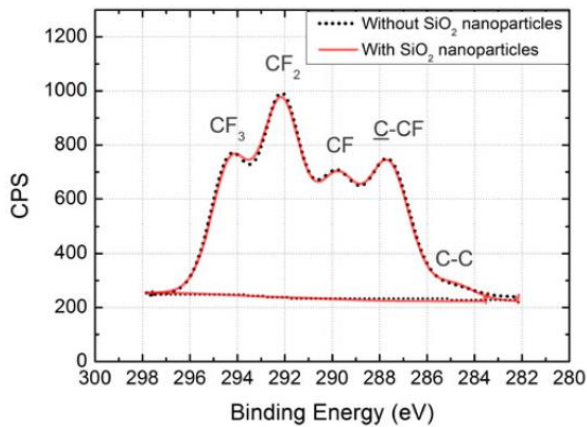

FIG. 13. HR C1s spectra of pp-$C_6F_{12}$ films in argon with $SiO_2$ nanoparticles (red curve) and without $SiO_2$ nanoparticles (black dashed curve). (0.2 L/min $C_6F_{12}$-carrier gas, 5 L/min carrier gas total, 50 W).

Roughness measurements of the films were performed using a stylus profilometer on a 900 µm² surface. The $R_{RMS}$ of the film synthesized in argon and helium plasma without any nanoparticles are 192 and 30 nm, respectively. The increase in the WCA with the addition of nanoparticles appears to be related to the rise in the $R_{RMS}$ value reaching up to 434 and 396 nm for argon and helium, respectively. The 5 and 12% WCA increase in argon and helium in presence of $SiO_2$ nanoparticles is explained by the thickness of the coatings. Indeed, as previously mentioned, the deposition rate of the fluorinated precursor in helium is much lower than in argon. This variation involves a thinner fluorinated covering layer for an equal deposition time. Since the nanoparticles are covered by a thinner layer in the case of helium, their effect on the roughness will be much more significant. Additional SEM images in Fig. 14 show that the film synthesized from helium is initially smooth and the nanoparticles are clearly visible, leading to a higher increase in the roughness and therefore in the wettability. Argon films are initially rough and the presence of nanoparticles has a weaker influence on the morphology and the resulting properties.

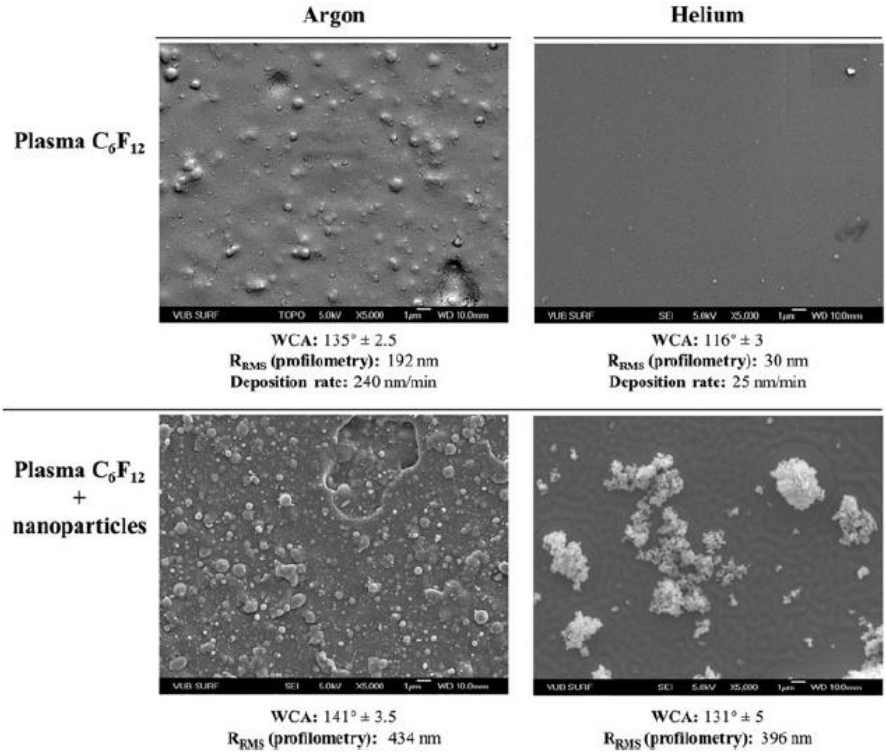

FIG. 14. SEM images (magnification 5000) and summary of the effect of the nanoparticles on the properties of coatings synthesized from argon and helium plasmas.

**3.3.2. Effect of nanoparticles: Synthesis of super-hydrophobic surfaces**

Super-hydrophobic films were synthesized by increasing the concentration of nanoparticles in the evaporated solutions. The influence of the concentration of nanoparticles was investigated by using 0.05, 0.5, 1, 2, and 5 g/L dispersions. Higher concentrations of nanoparticles lead to higher film roughness as shown in Fig. 15. The roughness RMS measured for a film of $C_6F_{12}$ synthesized by atmospheric plasma without any nanoparticles is of 192 nm while the presence of nanoparticles induces roughness of 434, 1503, and 3322 nm for concentrations of 0.05, 1, and 5 g/L, respectively.

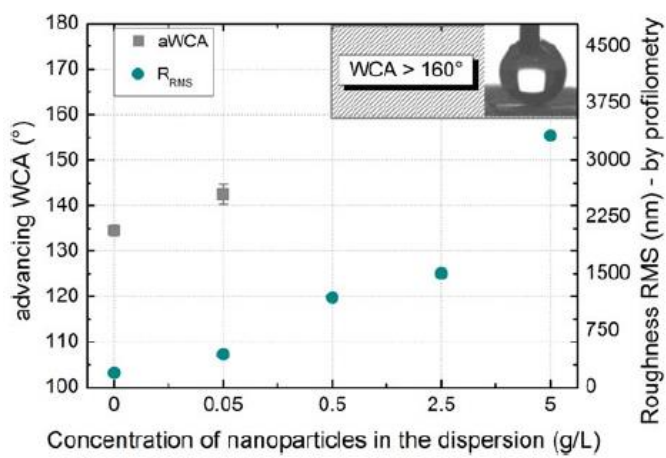

*FIG. 15. Advancing WCAs and roughness RMS as a function of the $SiO_2$ nanoparticles concentration for $C_6F_{12}$ coatings (Ar–$C_6F_{12}$ 50 W, 2 min).*

According to the wettability measurements, the roughness obtained with nanoparticles dispersions of 0.5 g/L or higher concentrations leads to a super-hydrophobic character of the surface. Indeed, for those plasma depositions, the advancing and receding WCAs are higher than 160° with a very low contact angle hysteresis (<5°). The surface is slippery since the water droplet does not adhere and roll across the surface of the sample as shown on Fig. 15. This result shows that the combination of rough surface morphology and low surface energy coatings required to obtain a super-hydrophobic surface is brought by the presence of nanoparticles and the low surface energy $C_xF_y$ coating, respectively.

The process of overcoating nanoparticles by fluorocarbon films at atmospheric pressure appears to be efficient to create super-hydrophobic surfaces. Kylian et al. reported the formation of slippery surfaces from a similar process but operated at low pressure through a magnetron sputtering of PTFE with lower deposition rates (50 nm/min versus 240 nm/min) than those presented in this article.55 This work shows that a simple and rapid process operated at atmospheric pressure, without constraints from vacuum systems and using only water as a solvent, can lead to promising results. Further studies using hydrocarbon precursors instead of fluorocarbon should be performed since $CH_2$ and $CH_3$ chemical groups are known to be hydrophobic. Indeed, Fanelli et al. showed that super-hydrophobic thin films could be obtained from aerosol-assisted atmospheric cold plasma deposition.89

## 4. Conclusion

The synthesis and texturization processes of fluorinated surfaces by means of atmospheric plasma were investigated and presented through an integrated study of both the plasma phase and the resulting material surface. Three methods producing super-hydrophobic surfaces and/or enhancing their hydrophobicity were evaluated:

(1) the modification of a fluorinated PTFE surface by the post-discharge of a RF plasma torch,
(2) the plasma deposition of fluorinated coatings by a DBD, and
(3) the incorporation of nanoparticles into those fluorinated films.

In the first approach, the addition of oxygen to the helium carrier gas showed a different behavior towards the surface properties. Oxygen is indeed required to enhance the hydrophobicity of the PTFE through its roughness increase but does not induce any change on the chemical composition. Atomic oxygen has been assumed to be responsible for this morphology modification since it etches preferably the amorphous phase; while the highly energetic helium metastables are supposed to be the main species responsible for the scission of –($CF_2$)– chains involving a higher etching without any preferential orientation.

The second approach focused on the deposition and intrinsic texturization of $C_xF_y$ films by PECVD in a DBD. We showed that the nature of the carrier gas is a crucial parameter since argon and helium plasmas induce different chemical composition, morphology and therefore hydrophobicity. Indeed, only argon could induce WCA higher than 115° by the creation of alveolar structure in the films while in helium, the films remain smooth. The characterization of the discharges allowed us to correlate the gas phase behavior and the surface properties since we identified a higher fragmentation in argon and the presence of oxygen in helium.

Finally, the results obtained from the PECVD of fluorocarbon precursors were compared with the hybrid coatings made of nanoparticles covered by a fluorinated film deposited by the DBD. The addition of nanoparticles induces an enhancement of the hydrophobicity through the roughness increase but does not alter the chemical composition of the film. Moreover, super-hydrophobic coatings characterized by the nonadherence of the water droplet at their surfaces were produced for high concentrations of nanoparticles.

# 5. Acknowledgments

This work was part of the I.A.P (Interuniversity Attraction Pole) programs "PSI _ Physical Chemistry of Plasma Surface Interactions—IAP-VII/12, P7/34—financially supported by the Belgian Federal Office for Science Policy (BELSPO). This work was also financially supported by the FNRS (Belgian National Fund for Scientific Research) for an "Aspirant Grant (FRFC Grant No. 2.4543.04)". Research in Mons is also supported by the Région Wallonne/European Commission FEDER program (SMARTFILM project) and FNRS-FRFC.

NOTE: THIS DOCUMENT IS A PRE-PRINT VERSION. YOU MAY USE IT AT YOUR OWN CONVENIENCE BUT ITS CONTENT MAY DEVIATE IN PLACES FROM THE FINAL PUBLISHED ARTICLE. FOR CITATION, REFER TO THE INFORMATION REPORTED IN THE INTRODUCTIVE TABLE30. R. d'Agostino, D.L. Flamm, and O. Auciello: Plasma Deposition, Treatment and Etching of Polymers (Academic Press, San Diego, USA, 1990).

31. F. Henry, F. Renaux, S. Coppée, R. Lazzaroni, N. Vandencasteele, F. Reniers, and R. Snyders: Synthesis of superhydrophobic PTFE like thin films by self-nanostructuration in a hybrid plasma process. Surf. Sci. 606, 1825 (2012).

32. Y-R. Wang, W-C. Ma, J-H. Lin, H-H. Lin, C-Y. Tsai, and C. Huang: Deposition of fluorocarbon film with 1,1,1,2- tetrafluoroethane pulsed plasma polymerization. Thin Solid Films 570, 445 (2014).

33. P. Favia, G. Cicala, A. Milella, F. Palumbo, P. Rossini, and R. d'Agostino: Deposition of super-hydrophobic fluorocarbon coatings in modulated RF glow discharges. Surf. Coat. Technol. 169–170, 609 (2003).

34. N.M. Mackie, N.F. Dalleska, D.G. Castner, and E.R. Fisher: Comparison of pulsed and continuous-wave deposition of thin films from saturated fluorocarbon/H2 inductively coupled rf plasmas. Chem. Mater. 9, 349 (1997).

35. R. d'Agostino, P. Favia, Y. Kawai, H. Ikegami, N. Sato, and F. Arefi-Khonsari: Advanced Plasma Technology (Wiley-VCH, Germany, 2008).

36. J. Hopkins and J.P.S. Babyal: Nonequilibrium glow discharge fluorination of polymer surfaces. J. Phys. Chem. 99, 4261 (1995).

37. M. Strobel, S. Corn, C.S. Lyons, and G.A. Korba: Plasma fluorination of polyolefins. J. Polym. Sci., Part A: Polym. Chem. 25, 1295 (1987).

38. F. Fanelli, F. Fracassi, and R. d'Agostino: Atmospheric pressure PECVD of fluorocarbon coatings from glow dielectric barrier discharges. Plasma Processes Polym. 4, S430 (2007).

39. F. Fanelli, F. Fracassi, and R. d'Agostino: Deposition and etching of fluorocarbon thin films in atmospheric pressure DBDs fed with Ar–CF4–H2 and Ar–CF4–O2 mixtures. Surf. Coat. Technol. 204, 1779 (2010).

40. I.P. Vinogradov and A. Lunk: Spectroscopic diagnostics of DBD in Ar/fluorocarbon mixtures—Correlation between plasma parameters and properties of deposited polymer films. Plasma Processes Polym. 2, 201 (2005).

41. I.P. Vinogradov, A. Dinkelmann, and A. Lunk: Deposition of fluorocarbon polymer films in a dielectric barrier discharge (DBD). Surf. Coat. Technol. 174–175, 509 (2003).

42. S.H. Kim, J-H. Kim, B-K. Kang, and H.S. Uhm: Superhydrophobic CFx coating via in-line atmospheric RF plasma of He─────────────CF4─────────────H2. Langmuir 21, 12213 (2005).

43. D. Liu, Y. Yin, D. Li, J. Niu, and Z. Fen: Surface modification of materials by dielectric barrier discharge deposition of fluorocarbon films. Thin Solid Films 517, 3656 (2009).

44. M. Nagai, O. Takai, and M. Hori: Atmospheric pressure fluorocarbon-particle plasma chemical vapor deposition for hydrophobic film coating. Jpn. J. Appl. Phys. 45, L460 (2006).

45. C-T. Hsieh, S-Y. Yang, and J-Y. Lin: Electrochemical deposition and superhydrophobic behavior of ZnO nanorod arrays. Thin Solid Films 518, 4884 (2010).

46. J. Wang, A. Li, H. Chen, and D. Chen: Synthesis of biomimetic superhydrophobic surface through electrochemical deposition on porous alumina. J. Bionic Eng. 8, 122 (2011).

47. B. Balu, J.S. Kim, V. Breedveld, and D.W. Hess: Tunability of the adhesion of water drops on a superhydrophobic paper surface via selective plasma etching. J. Adhes. Sci. Technol. 23, 361 (2009).

48. J.C. Shearer, M.J. Fisher, D. Hoogland, and E.R. Fisher: Composite SiO2/TiO2 and amine polymer/TiO2 nanoparticles produced using plasma-enhanced chemical vapor deposition. Appl. Surf. Sci. 256, 2081 (2010).

49. R.V. Lakshmi, T. Bharathidasan, P. Pera, and B.J. Basu: Fabrication of superhydrophobic and oleophobic sol–gel nanocomposite coating. Surf. Coat. Technol. 206, 3888 (2012).
18

Journal of Materials Research, 2015, Vol. 30, Issue 21, 3177-3191, http://dx.doi.org/10.1557/jmr.2015.279